\documentclass[%
 reprint,
 amsmath,amssymb,
 aps,prl
]{revtex4-1}

\usepackage{graphicx}
\usepackage{dcolumn}
\usepackage{bm}
\usepackage{hyperref}

\usepackage[separate-uncertainty=true,multi-part-units=single]{siunitx}

\usepackage{color}

\begin{document}

\preprint{APS/123-QED}

\title{Nonlinear Electrophoresis of Highly Charged Nonpolarizable Particles}

\author{Soichiro Tottori}
\affiliation{ 
Cavendish Laboratory, Department of Physics, University of Cambridge, JJ Thomson Avenue, Cambridge, CB3 0HE,
United Kingdom
}
\author{Douwe Jan Bonthuis}
\affiliation{
Institute of Theoretical and Computational Physics, Graz University of Technology, 8010 Graz, Austria
}
\author{Karolis Misiunas}%
\affiliation{ 
Cavendish Laboratory, Department of Physics, University of Cambridge, JJ Thomson Avenue, Cambridge, CB3 0HE,
United Kingdom
}
\author{Ulrich F. Keyser}\email{ufk20@cam.ac.uk}
\affiliation{ 
Cavendish Laboratory, Department of Physics, University of Cambridge, JJ Thomson Avenue, Cambridge, CB3 0HE,
United Kingdom
}

\date{\today}

\begin{abstract}
Nonlinear field dependence of electrophoresis in high fields has been investigated theoretically, yet experimental studies have failed to reach consensus on the effect.
In this work, we present a systematic study on the nonlinear electrophoresis of highly charged submicron particles in applied electric fields of up to several \SI{}{\kilo\volt/\centi\meter}. 
First, the particles are characterized in the low-field regime at different salt concentrations and the surface charge density is estimated. Subsequently, we use microfluidic channels and video tracking to systematically characterize the nonlinear response over a range of field strengths.
Using velocity measurements on the single particle level, we prove that nonlinear effects are present at electric fields and surface charge densities that are accessible in practical conditions.
Finally, we show that nonlinear behavior leads to unexpected particle trapping in channels.
\end{abstract}

\maketitle

\begin{figure}
\includegraphics{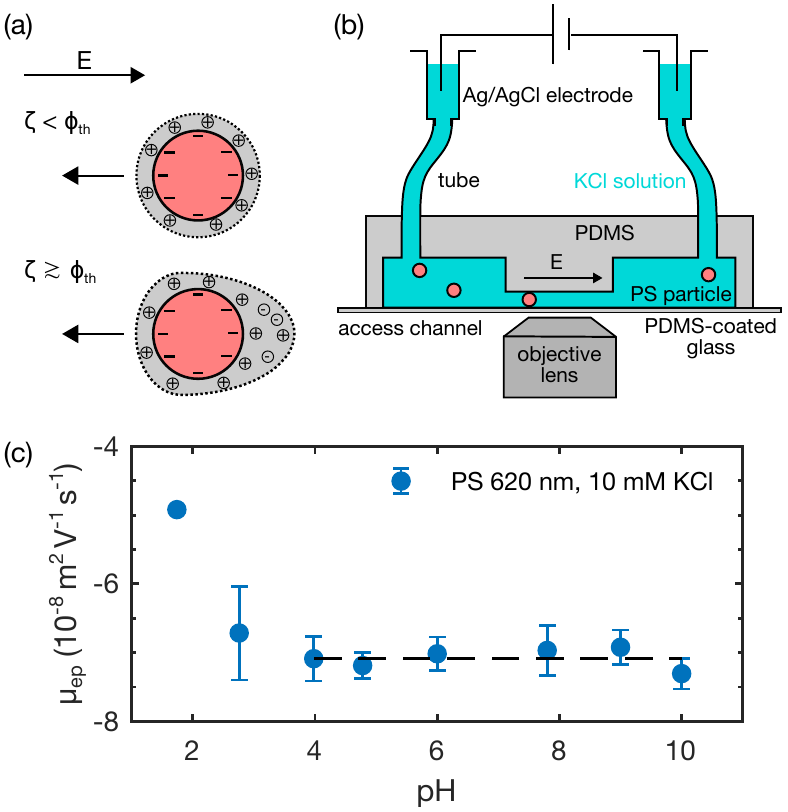}
\caption{\label{fig:1}(a) Schematic illustrations of the counterion cloud around particles with low and high zeta potentials under an electric field $E$. 
(b) Schematic illustration of our experimental system. 
Microfluidic chip is connected to two open reservoirs. 
Particle motion in channels is imaged using a camera via the objective.
(c) Bulk electrophoretic mobility of PS particle (\SI{620}{\nano\meter} in diameter) as a function of \textit{p}H. 
Dotted line is the averaged value of $\mu_{\textrm{ep}}$ for \textit{p}H $\geq 4$.}
\end{figure}

Electrophoresis has been widely used for sensing, filtration, manipulation, and the separation of molecules and particles, particularly recently using micro- and nanofluidic devices~\cite{Yukimoto2013, Liu2014, Wu2014, Angeli2015}. 
On these length scales, the applied electric field may reach $\sim$kV/cm because the confinement focuses the electric field,  even with a moderate applied voltage to the system.
The electrophoretic velocity $v_{\textrm{ep}}$ of a rigid nonpolarizable sphere in the limit of low zeta potential $\zeta$ and low applied electric field $E$ is
\begin{equation}
\label{eq:1}
v_{\textrm{ep}} = \frac{2\epsilon }{3\eta} f(\kappa a) \zeta E,
\end{equation}
where $\epsilon$ and $\eta$ are the permittivity and viscosity of the fluid, respectively~\cite{Masliyah2006}.
$f(\kappa a)$ is Henry's function, where $\kappa^{-1}$ and $a$ are the Debye length and particle radius, respectively~\cite{Henry1931, Dukhin1993}.
Within a canonical model (homogeneous $\epsilon$ and $\eta$, no specific ion-surface interactions, no hydrodynamic slip) and using a low zeta potential ($\left|\zeta\right| < \phi_{\textrm{th}} = k_B T /e \approx\SI{26}{\milli\volt}$), $\zeta$ calculated using Eq.~(\ref{eq:1}) corresponds to the homogeneous electrostatic potential at the particle surface.
In real systems, however, the zeta potentials of particles may exceed $\phi_{\textrm{th}}$, in which case the direct action of the electric field on the double layer as well as the advection of ions by the flow result in an asymmetric shape of both coion and counterion clouds, as schematically shown in Fig.~\ref{fig:1}(a)~\cite{Obrien1978, Grosse2015}.
For intermediate values of $\kappa a$, the double-layer distortion, known as relaxation effect, leads to a nonlinear dependence of $v_{\textrm{ep}}$ on $\zeta$~\cite{Obrien1978, Dukhin1993, Squires2016}.
Therefore, in case $\left|\zeta\right| \gtrsim \phi_{\textrm{th}}$, the model including relaxation effect \cite{OBrien1983, Chen1992} needs to be used instead of Eq.~(\ref{eq:1}) to extract the surface potential that reproduces $v_{\textrm{ep}}$ within the canonical model.

In addition, the relaxation effect is field dependent, complicating the interpretation of experimental results even further (see Supplemental Material~\cite{SM}).
This nonlinear effect, which is different from the nonlinear electrokinetics found for polarizable particles \cite{Bazant2004}, is not well understood for intermediate to large $\kappa a$ in contrast to the case of small $\kappa a$ in nonpolar electrolytes~\cite{Stotz1978, Thomas2008, Gacek2012, Guo2013, Strubbe2013, Khair2018}.
Theoretical analyses have predicted that the electrophoretic velocity becomes nonlinear to the applied field at a moderate to high field $\beta = a E /\phi_{\textrm{th}} \gtrsim 1$~\cite{Shilov2003, Schnitzer2013, Schnitzer2014, Sherwood2018}.
However, a well-controlled experiment in this regime has been lacking. Indeed, there are experimental reports showing that particle electrophoretic velocity is linear~\cite{Kumar2006} and nonlinear~\cite{Shilov2003} under similar conditions.
Even for the nonlinear case, the properties of particles were not clearly characterized.
Moreover, the zeta potential was reported to be severalfold lower than the value used for fitting one of the models, and the difference was then attributed to the divergence of the slip plane from the actual particle surface~\cite{Mishchuk2010, Mishchuk2011}. Here, we resolve the conflicting results found in the literature by systematic experiments in combination with detailed simulations.

We used high-speed video tracking to study the nonlinear electrophoresis of highly charged particles in microfluidic channels.
We characterized particle mobility with different salt concentration, and subsequently measured the nonlinear electrophoretic velocity under high field up to $\beta \approx 3$.
The experimental data were compared with the values obtained with coupled Stokes-Poisson-Nernst-Planck (SPNP) simulations, showing that high field particle electrophoresis experiments can be quantitatively explained with our SPNP model.
We also demonstrate that particles can be trapped in the channel at high fields due to this nonlinear effect in combination with the electro-osmotic flow resulting from the channel wall. 

Microfluidic channels were made out of polydimethylsiloxane (PDMS) cast onto microfabricated molds~\cite{Pagliara2011}.
The mold for the ``wide'' channel ($w = \SI{12.7}{\micro\metre}$, $h = \SI{5.2}{\micro\metre}$, $l= \SI{100}{\micro\metre}$) was fabricated with SU-8 photoresist.
The ``narrow'' channels ($w = \SI{750}{\nano\metre}$, $h = \SI{750}{\nano\metre}$, $l = \SI{10}{\micro\metre}$) were made by depositing platinum onto silicon substrates with focused ion beam.
The access channels for wide and narrow channels were made out of SU-8 ($h = \SI{100}{\micro\metre}$) and AZ 9260 ($h = \SI{12}{\micro\metre}$), respectively.
The bottom glass slides were coated with a thin layer of PDMS, then plasma bonded to the PDMS channels~\cite{Deshpande2016}.
The overview of our experimental system is shown in Fig.~\ref{fig:1}(b).
The channels were connected to the two reservoirs through access channels and tubes.
The reservoir height was adjusted to eliminate the pressure difference between both sides before each experiment.
Voltage ($\leq$ \SI{35}{\volt}) was applied through Ag/AgCl electrodes connected to a source measure unit (Keithley 2450).
The particle motion was recorded at maximum \SI{2000}{fps} and tracked by custom-written software after experiments.
Conversion coefficients $k$ from voltage to field inside a channel, $E=kV$, were calibrated for each microfluidic chip design by measuring the current $I$ through the system, as $k = I/(\rho AV)$, where $\rho$ and $A$ are the conductivity of the electrolyte and total channel cross sectional area, respectively.

The particles used in our experiments were polystyrene (PS, from Polysciences and Bangs Labs) and poly(methyl methacrylate) (PMMA, from Microparticles GmbH). 	Bulk electrophoretic mobilities were measured with Laser Doppler Microelectrophoresis (Zetasizer Nano ZSP, Malvern). All particles were diluted in KCl solution to the final concentration (volume fraction) of approximately $2.5 \times 10^{-5}$, $1\times10^{-5}$, and $2.5$--$10 \times 10^{-4}$ for narrow and wide microchannel experiments and Laser Doppler Microelectrophoresis measurement, respectively. 
In Fig.~\ref{fig:1}(c), the electrophoretic mobility of PS particles in \SI{10}{\milli M} KCl is plotted as a function of \textit{p}H. 
The \textit{p}H was adjusted by adding small quantities of KOH or HCl. 
The observed negative mobility of \SI{\approx -7}{\times 10^{-8}\meter^2\volt^{-1}\second^{-1}} is due the negative surface charge associated with the sulfate end groups on PS, originating from the decomposition of the initiator used during particle synthesis (initiator: K$_2$S$_2$O$_8$, end group: -OSO${_3}^-$K$^+$). 
The reported values vary but typically are in the range of several tens of mC/m$^2$. 
As the dissociation constant $pK_a$ of sulfate end groups is below 2~\cite{Elimelech1990, Midmore1996, Behrens2001}, no significant change in electrophoretic mobility was observed for \textit{p}H $\geq $ 4. 
We decided that no buffer was necessary in the following experiments, and \textit{p}H values were all kept in the range of 6--9.

Coupled SPNP simulations were performed using a finite element method (COMSOL Multiphysics ver.~4.4).
The particle was fixed in a large enough space ($=100a$) with open boundary conditions. The surface charge density was set at a constant value on the particle surface.
The electro-osmotic flow generated far away from the particle was used to calculate the electrophoretic mobility of the particle~\cite{Masliyah2006}. More details on the simulations can be found in Supplemental Material~\cite{SM}.

In Fig.~\ref{fig:2}, the electrophoretic mobilities $\mu_\textrm{ep}$ of various particles measured with Laser Doppler Micro-electrophoresis are plotted as a function of KCl concentration $c$. The field used in the measurement was below \SI{0.025}{\kilo\volt/\centi\metre} ($\beta \approx 0.05$); thus nonlinear effects were negligible. The electrophoretic mobility curves of PS particles display negative peaks at about \SI{10}{\milli M}, in a good agreement to previous literature values for other highly charged particles~\cite{Elimelech1990, Antonietti1997, Borkovec2000, Kobayashi2014}. The negative peak of PMMA particles, which had lower charge density, was found at around \SI{1}{\milli M}.

In order to estimate the surface charge density from these data, we fitted the approximate model including the relaxation effect by Chen and Keh~\cite{Chen1992} (valid for $\kappa a \gtrsim 20$) to the experimental data  with a surface charge density as the only fitting parameter (see Supplemental Material for more details~\cite{SM}).
The extracted surface charge densities of PS \SI{620}{\nano\meter}, PS \SI{370}{\nano\meter}, PMMA \SI{520}{\nano\meter}, and PS-COOH \SI{380}{\nano\meter} are $-51.2$, $-41.7$, $-13.1$, and \SI{-11.2}{\milli\coulomb/\meter^2}, respectively. 
Note that these surface charge densities are not necessarily the bare surface charge densities because we do not take into account the details of the surface, such as local viscosity and permittivity.
However, since these values are employed in the numerical SPNP simulations to predict the magnitude of the nonlinearity using the canonical model of electrophoresis, this treatment is fully consistent.

\begin{figure}
\includegraphics{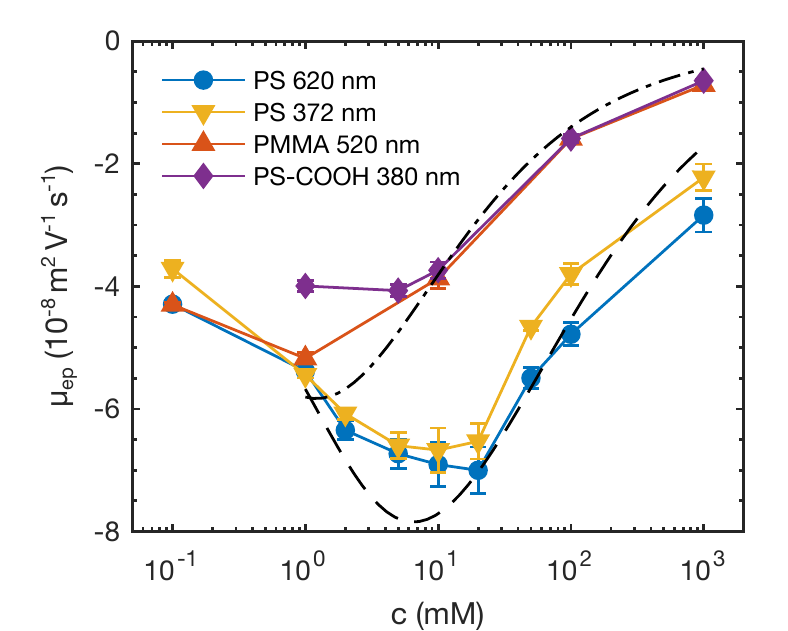}
\caption{\label{fig:2}Bulk electrophoretic mobility as a function of salt concentration. 
Dashed and dash-dotted lines are the fitted curves to the Chen and Keh model~\cite{Chen1992} for PS \SI{620}{\nano\meter} and PMMA \SI{520}{\nano\meter}, respectively. 
Estimated surface charge densities of PS \SI{620}{\nano\meter}, PS \SI{370}{\nano\meter}, PMMA \SI{520}{\nano\meter}, and PS-COOH \SI{380}{\nano\meter} are $-51.2$, $-41.7$, $-13.1$, and $\SI{-11.2}{\milli\coulomb/\meter^2}$, respectively.}
\end{figure}

Having established an estimate for the surface charge density, the particle velocity in high fields was characterized using a microfluidic channel. 
Hydrodynamic and electric wall effects were negligible in this wide channel since the channel height and width were large enough~\cite{Chiu2016}. 
The two types of particles (PS \SI{620}{\nano\meter} and PMMA \SI{520}{\nano\meter})  were suspended in \SI{1}{\milli M} KCl. 
Previously, two methods have been proposed to extract particle mobility in a high field: using asymmetric oscillating fields~\cite{Shilov2003, Barany2009, Mishchuk2010, Mishchuk2011} and measuring bulk flow rate from the hydraulic height in a reservoir~\cite{Kumar2006, Youssefi2016}. 
Here we used microfluidic channels and dc voltage to directly monitor particle motions, which allowed us to acquire nonlinear electrophoretic velocity on the single particle level precisely, without transient behavior, bubble generation, and pressure flow. 
	 
\begin{figure}
\includegraphics{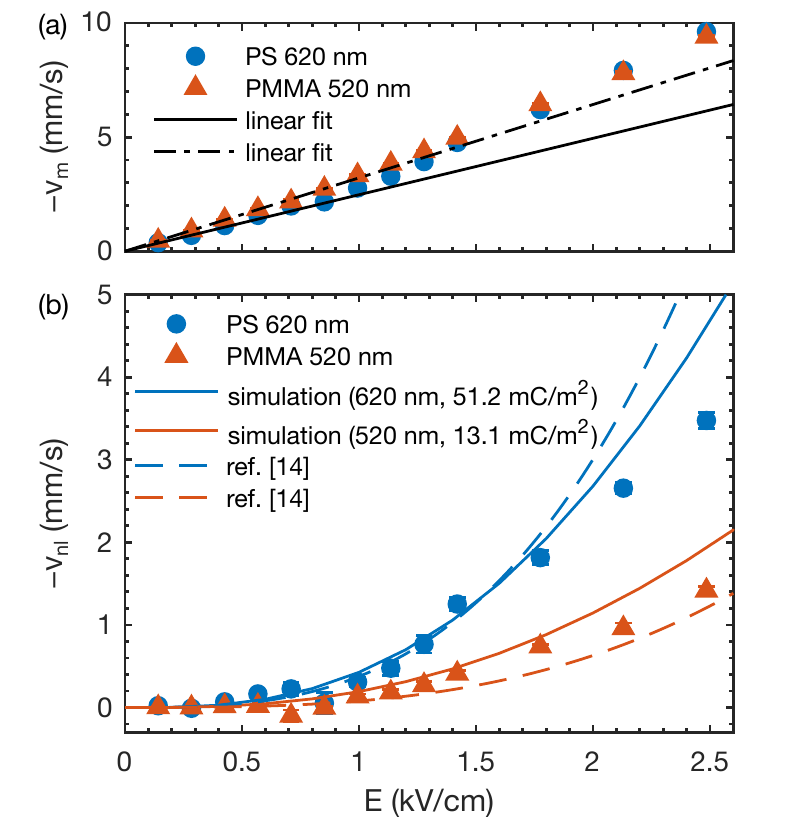}
\caption{\label{fig:3}(a) Measured particle velocity $v_{m}$ ($=v_{\textrm{ep}} +v_{\textrm{eof}}$) as a function of the applied field in the wide channel ($c = \SI{1}{\milli M}$).
Solid and dashed lines are linear fits of the first two points.
(b) Nonlinear velocity components $v_{\textrm{nl}}$ as a function of the applied field.
Solid lines are the SPNP simulation results.
Error bars in (a) and (b) are both standard errors of the mean and within the symbol size.
Dashed lines are the cubic term from the model by Schnitzer and Yariv in~\cite{Schnitzer2014}.}
\end{figure}

In Fig.~\ref{fig:3}(a), the measured particle velocity $v_{m}$ ($=v_{\textrm{ep}}+v_{\textrm{eof}}$, the sum of the electrophoretic and electro-osmotic velocities) is plotted as a function of an applied field with the maximum field strength \SI{\approx 2.5}{\kilo\volt/\centi\metre}, which is equivalent to $\beta \approx 3$ (for $2a = \SI{620}{\nano\meter}$). 
The solid and dashed lines are the linear fits to the first two points of each data set. 
Clearly, the particle velocities increased superlinearly with the applied field for both types of particles.
Figure~\ref{fig:3}(b) shows the nonlinear components of the particle velocity $v_{\textrm{nl}}$, which was extracted by subtracting the linear components obtained from linear fits shown in Fig.~\ref{fig:3}a. This process is important because it is difficult to measure $v_{\textrm{eof}}$ precisely \textit{in situ}.
The solid lines are the nonlinear component of the SPNP simulation results with the actual size of the particle and the surface charge densities of \SI{-51.2} and \SI{-13.1}{\milli\coulomb/\meter^2}, respectively. The simulations showed a good agreement with the experiments. This implies that we can predict the magnitude of nonlinearity well with the estimated surface charge density.

Our results contradict the earlier report by Kumar \textit{et al.}~\cite{Kumar2006}, where they concluded that the electrophoretic velocity was always linear. Our results are qualitatively in line with the measurements performed by Shilov \textit{et al.}~\cite{Shilov2003} and Barany~\cite{Barany2009} using oscillating fields. 
However, a direct comparison is difficult because neither the zeta potential nor surface charge density of their particles were reported. 
Instead, the model in \cite{Shilov2003} was fitted to the experimental data with zeta potential as a fitting parameter. 
Mishchuk~\textit{et al.} used the same approach and reported that their measured particle zeta potential (\SI{-28}{\milli\volt}) was significantly lower than the zeta potential estimated using Shilov's model (\SI{-175}{\milli\volt})~\cite{Shilov2003}, and attributed this difference to the divergence of the slip plane and particle surface~\cite{Mishchuk2010, Mishchuk2011}. 
However, the zeta potential was estimated using Eq.~(\ref{eq:1}) even at low salt concentrations ($c =$ \SI{0.1}{\milli M}). Thus, it is likely that this value was underestimated compared to the actual zeta potential. 
In contrast, our results suggest that nonlinear effects can be accurately predicted by taking into account the relaxation effect with extracted surface charge densities.

For the analytical approach in the thin double layer limit and small P\'eclet number, the first two allowed terms by symmetry are described as $v_{\textrm{ep}} = \mu_{\textrm{ep}}E + \mu_{\textrm{ep}}^{(3)} E^3$, where both coefficients $\mu_{\textrm{ep}}$ and $\mu_{\textrm{ep}}^{(3)}$ are independent of $E$~\cite{Schnitzer2013, Schnitzer2014}. 
The coefficients $\mu_{\textrm{ep}}^{(3)}$ obtained from our results (PS \SI{620}{\nano\meter} and PMMA \SI{520}{\nano\meter}) in Fig.~\ref{fig:3}(b) were approximately $\SI{-21.4}{}(\pm 2.2)$ and $\SI{-9.9}{}(\pm 1.0)$ \SI{}{\times 10^{-20} \metre^4 \second^{-1}\volt^{-3}}, respectively.
Based on the model proposed by Schnitzer \textit{et al.}~\cite{Schnitzer2014, SM} for small Dukhin and P\'eclet numbers, the nonlinear coefficients for these parameters are $\SI{-37.6}{}$ and $\SI{-7.9}{\times 10^{-20} \metre^4 \second^{-1}\volt^{-3}}$, as also shown in Fig.~\ref{fig:3}(b).
Interestingly, although the model is based on the assumption of small Dukhin and P\'eclet numbers, it still shows a reasonable agreement to our experimental data $\mu_{\textrm{ep}}^{(3)}$.
These results imply that the two particles with similar $\mu_{\textrm{ep}}$ can be still separated by the difference of $\mu_{\textrm{ep}}^{(3)}$.

\begin{figure}
\includegraphics{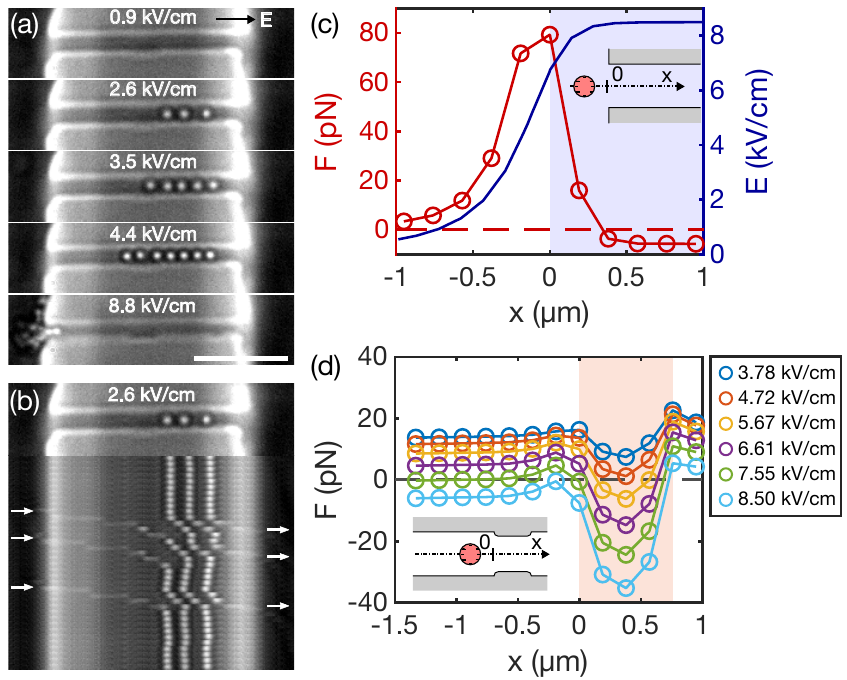}
\caption{\label{fig:4}Electric field induced trapping and transport.
(a) Images show 0, 3, 5, and 7 particles (PS-COOH \SI{380}{\nano\meter}) trapped inside the channel for $E = 0.9, 2.6, 3.5$, and \SI{4.4}{\kilo\volt/\centi\metre}, respectively. 
No particle entered the channel at \SI{8.8}{\kilo\volt/\centi\metre}.
Scale bar equals \SI{5}{\micro\meter}.
(b) Example of transport mode for a channel with three particles at \SI{2.6}{\kilo\volt/\centi\metre}.
Spatiotemporal diagram displaying the cascading motion of trapped particles. 
During the shown time frame of \SI{72}{\milli\second} three particles entered from the left (arrows) releasing the particles at the right end (arrows).
(c) Simulated force landscape near the entrance (left) and local external electric field (right).
Shadowed area corresponds to the inside of the channel.
(d) Simulated force landscape near the constriction.
Shadowed area corresponds to the constriction.
Field strengths are the values at the center of an empty channel.}
\end{figure}

Last, we discovered that this nonlinear electric field dependence allows for voltage-controlled particle trapping in confinement. 
Here, slightly lower charged particles (PS-COOH \SI{380}{\nano\meter}) in \SI{5}{\milli M} KCl were introduced into PDMS-glass-bonded narrow channels. 
The electro-osmotic flow was stronger than the electrophoresis of the particles; thus the particles were always transported by the electro-osmotic flow at low fields. 
The mobility of particle inside the channel $\mu_c$ ($=\mu_\mathrm{ep}+\mu_\mathrm{eof}$) was \SI{\approx 0.8}{\times 10^{-8} \metre^2 \volt^{-1}\second^{-1}} at $E \approx$ \SI{0.088}{\kilo\volt/\centi\metre} ($\beta \approx 0.065$). 
Interestingly, at a sufficiently high field ($E \gtrsim$ \SI{2.6}{\kilo\volt/\centi\metre} or $\beta \gtrsim 1.9$), the particles were trapped at the inside or entrance of the channel, as shown in Fig.~\ref{fig:4}(a). 
The number of trapped particles increased with the applied field from 3, 5, 7 for 2.6, 3.5, and \SI{4.4}{\kilo\volt/\centi\metre} ($\beta \approx$ 1.9, 2.6, and 3.3), respectively.
No particle entered the channel at $\gtrapprox \SI{8.8}{\kilo\volt/\centi\metre}$.
When trapping occurred inside the channel, trap-and-release cascade motion was observed: when one additional particle was entering, the initially trapped particle was released. 
The spatiotemporal diagram of this cascade motion is shown in Fig.~\ref{fig:4}b. 

In order to understand this trapping mechanism, we numerically simulated the force landscape near the entrance of the channel, as shown in Fig.~\ref{fig:4}(c).
Force was calculated as the sum of electrostatic and hydrodynamic forces on the fixed particle ($2a =\SI{380}{\nano\meter}, \sigma_p =\SI{-11.2}{\milli\coulomb/\meter^2}$). 
The surface charge density of the channel wall was estimated to be $\sigma_w = \SI{-12.7}{\milli\coulomb/\meter^2}$ from the particle velocity inside the channel at low fields.
The confinement ratio was set $a/R =0.5$.
The force landscape turns from positive to negative near the entrance because of the sharp increase of the field.
We further investigated the effect of small inhomogeneity on force landscapes by adding a small constriction in the channel (\SI{20}{\nano\meter} $\approx 0.05R$ in height, $R$ in length) in Fig.~\ref{fig:4}(d).
At intermediate fields ($5.67$, $6.61$, and \SI{7.55}{\kilo\volt/\centi\metre}), force reversal was observed near the constriction inside the channel, indicating that particles can be trapped inside the channel.
The force landscapes decrease with increasing fields, resulting in larger trapping forces.
The larger trapping forces allow for a larger number of trapped particles, before the outermost particle is pushed out of the trapping site by steric or hydrodynamic interaction with an incoming particle. 
At a sufficiently high field (\SI{8.50}{\kilo\volt/\centi\metre}), no particle can be trapped inside the channel since the force landscape is negative at the entrance, preventing particles from entering the channel.
Clearly, the trapping strength is dependent on the characteristics of the constrictions, such as height and length, yet our numerical results capture the trapping and cascade motions observed in experiments.
This trapping effect due to nonlinear field dependence of electrophoresis is intriguing, as it is clearly different in origin from well characterized dielectrophoretic trapping~\cite{Kang2006}.

We experimentally proved that highly charged nonconducting particles exhibit nonlinear electrophoretic velocity at high fields.
Our results include measurements ranging from the low field to the high field limit spanning several orders of magnitude in salt concentration.
Our high-speed video tracking results allowed for extracting the particle mobility in channels and comparing the results with the numerical models. We found a near quantitative agreement with numerical simulations using finite element methods. Finally we showed that the nonlinear mobility can give rise to unexpected particle trapping that is controlled by the applied field. Our results are relevant for various fluidic systems and processes, including filtration, separation, and manipulation.

\begin{acknowledgments}
We thank Dr. Ory Schnitzer for his suggestion and Dr. Alice Thorneywork for her comments on the manuscript.
We also thank Bangs Lab and microParticles GmbH for providing us the information on the formation of microparticles. S.T. is supported by Nakajima Foundation Scholarship and John Lawrence Cambridge Trust International Scholarship. K.M. and U.F.K. acknowledge funding from an ERC consolidator grant (DesignerPores 647144).
\end{acknowledgments}



%

\end{document}



\title{Supplemental Material:\\Nonlinear Electrophoresis of Highly Charged Nonpolarizable Particles}

\author{Soichiro Tottori}
\affiliation{ 
Cavendish Laboratory, Department of Physics, University of Cambridge, JJ Thomson Avenue, Cambridge, CB3 0HE,
United Kingdom
}
\author{Douwe Jan Bonthuis}
\affiliation{%
Institute of Theoretical and Computational Physics, Graz University of Technology, 8010 Graz, Austria
}%
\author{Karolis Misiunas}%
\affiliation{ 
Cavendish Laboratory, Department of Physics, University of Cambridge, JJ Thomson Avenue, Cambridge, CB3 0HE,
United Kingdom
}

\author{Ulrich F. Keyser}\email{ufk20@cam.ac.uk}
\affiliation{ 
Cavendish Laboratory, Department of Physics, University of Cambridge, JJ Thomson Avenue, Cambridge, CB3 0HE,
United Kingdom
}

\date{\today}


\maketitle
\section{Nonlinear terms in particle electrophoresis}

The canonical model of electrophoresis consists of a spherical particle with constant surface potential $\zeta$, in a medium with a constant dielectric constant $\epsilon$, constant viscosity $\eta$, only electrostatic (no specific) ion-surface interactions and a no-slip hydrodynamic boundary condition on the surface of the particle.
Using this model in the limit of low $\zeta$ and low applied electric field $E$, the electrophoretic mobility of a spherical particle is a linear function of $\zeta$ and $E$.
The dependence on the particle size $a$ and the inverse Debye length $\kappa$ is given by the Henry function $f(\kappa a)$,
\begin{equation}
v_{\textrm{ep}} = \frac{2\epsilon \zeta}{3 \eta} f(\kappa a) E.
\label{eq:NT1}
\end{equation}
For small values of $\kappa a$ the Henry function approaches unity, whereas for $\kappa a \rightarrow \infty$ it approaches $3/2$.
At large values of $\zeta$ and intermediate values of $\kappa a$, the mobility becomes a nonlinear function of $\zeta$ \cite{Wiersema1966, Dukhin1993}.
In contrast, at both small and large values of $\kappa a$, the electrophoretic velocity remains a linear function of $\zeta$ for all values of $\zeta$.

Studying the interfacial layer in molecular detail shows that
the electrophoretic mobility of a spherical particle depends on its surface charge density $\sigma$ in a complex way, depending on the inhomogeneous viscosity and the dielectric properties of the interfacial layer, as well as on specific ion-surface interactions and hydrodynamic slip~\cite{Bonthuis2012}.
Therefore, Eq.~(\ref{eq:NT1}) effectively provides only a definition of the $\zeta$ potential as an effective potential that reproduces the particle's electrophoretic mobility within the canonical model described above.
Instead of defining a $\zeta$ potential and using Eq.~(\ref{eq:NT1}), we can thus define the linear electrophoretic mobility $\mu_{\textrm{ep}}(\epsilon,\eta,\kappa a,\sigma)$ and write
\begin{equation}
v_{\textrm{ep}} = \mu_{\textrm{ep}} E,
\label{eq:NT2}
\end{equation}
where the only necessary approximation is that the electric field is small: $aEe \ll k_BT$~\cite{OBrien1978}.
Clearly, when extracting $\zeta$ from $\mu_{\textrm{ep}}$ at intermediate values of $\kappa a$ and large values of the surface charge density with the aim of using $\zeta$ in a numerical solution of the canonical model, it should be kept in mind that a more accurate relation should be used instead of Eq.~(\ref{eq:NT1})~\cite{Wiersema1966},
such as the one employed in the main text~\cite{Chen1992}.

This situation changes at high electric field strengths, when the nonlinear terms in the Stokes-Nernst-Planck equations become non-negligible.
The magnitudes of the nonlinear terms depend on the surface charge density, as well as on the mobility of the particle and the ions.
Here we give a brief explanation of the source of the nonlinearity.
We start with the coupled Stokes and Nernst-Planck equations for a single ionic species,
\begin{equation}
\label{eq:NT3}
\vec{J} = -D \left[\nabla c +zc\nabla\psi^{*}\right] + \vec{u} c,
\end{equation}
with $\psi^{*}$ being the electrostatic potential normalized with respect to the thermal voltage $\psi^{*} = \psi/\phi_{\textrm{th}} $, $c$ being the concentration of one type of ion species, $\vec{J}$ and $\vec{u}$ being the ion flux and fluid velocity, respectively, and $D$ and $z$ being the diffusion coefficient and the valence of the ionic species. In the steady state,
\begin{equation}
\label{eq:NT4}
0= \nabla \cdot \left[ D\nabla c +zDc\nabla\psi^{*} - \vec{u} c\right].
\end{equation}
When an electric field is applied, the system departs from equilibrium. The electric field strength $E = -|\nabla \psi|_{\infty}$ defines a length scale at which the change of the potential is equal to the thermal voltage $\phi_{\textrm{th}}$. To normalize the equations, this length scale is compared to the radius $a$ of the particle, giving rise to the definition of the non-dimensional electric field  $\beta = aE/\phi_{\textrm{th}}$. Now we split the ion concentration and the potential into their equilibrium values, $c_0$ and $\psi^*_0$, and the perturbations due to the applied electric field, $\delta c$ and $\delta \psi^*$. In equilibrium, $\vec{u} = 0$, and therefore $\nabla \cdot \left[ D\nabla c_0 +zDc_0\nabla\psi^{*}_0\right] = 0$, leading to
\begin{equation}
\label{eq:NT5}
0= \nabla \cdot \left[ D \nabla \delta c +zDc_0\nabla\delta\psi^{*} +zD\delta c\nabla\psi^{*}_0 - \vec{u} c_0 + zD\delta c\nabla\delta\psi^{*} -\vec{u}\delta c\right].
\end{equation}
To leading order, $\delta c$, $\delta \psi^*$ and $\vec{u}$ are proportional to $\beta$. 
O'Brien and White employed a first order approximation with respect to the field $\beta$, neglecting the second-order terms (the final two terms) of Eq.~(\ref{eq:NT5})~\cite{OBrien1978}.
This procedure leads to an electrophoretic mobility which is a linear function of the applied electric field for all values of the surface charge density.
Note that the extra flux due to the action of the equilibrium field on the perturbation of the double layer and the gradient of the double layer perturbation (both parts of the surface conductance) are taken into account in the linearized equations, and thus do not produce a nonlinear electrophoretic velocity \textit{per se}, regardless of the magnitude of the surface potential.
Therefore, any nonlinear effect must be produced by the final two terms of Eq.~(\ref{eq:NT5}), which have in common that they are proportional to  $\delta c$, \textit{i.e.} to the distortion of the electric double layer upon application of the electric field.

Rearranging Eq.~(\ref{eq:NT5}) gives
\begin{equation}
\label{eq:NT6}
0 = \nabla \cdot \left[ a \nabla \delta c + az c_0 \nabla\delta\psi^{*} +azD\delta c\nabla\psi^{*}_0 - \frac{a \vec{u} c_0}{D} + \left(az\nabla\delta\psi^{*} -\frac{a\vec{u}}{D}\right) \delta c \right].
\end{equation}
For an electric field which is large on the scale of the particle, \textit{i.e.} $\beta \geq 1$, the final two terms of Eq.~(\ref{eq:NT6}) cannot be ignored. 
The concentration perturbation $\delta c$ depends linearly on the applied electric field and on the ion concentration in the electric double layer~\cite{OBrien1978}.
Therefore $\delta c$ will be larger for large surface charge densities,
but
the relation between $\delta c$, the bare surface charge density and the $\zeta$ potential estimated from mobility measurements is intricate and depends strongly on the interfacial dielectric environment, the interfacial viscosity and the specific ion-surface interactions.
In the thin double-layer limit, the criterion for negligible concentration polarization is often expressed in terms of the Bikerman or Dukhin number at high values of $\zeta$, $\textrm{Du} = \exp{|\zeta|/2}/(\kappa a)$ \cite{Dukhin1991}.
In the limit of vanishing $\textrm{Du}$, $v_{\textrm{ep}}$ becomes a linear function of the electric field for any field strength, but as the present experiments show, a sizable nonlinear component is observed already for Dukhin numbers significantly smaller than 1 ($\textrm{Du} \approx 0.4$ and $\textrm{Du} \approx 0.1$ for the PS \SI{620}{\nano\meter} and PMMA \SI{520}{\nano\meter} particles, respectively).
For the first term inside the curved brackets, we find $a\nabla\delta\psi^{*}/\beta \sim 1$, and therefore this term cannot be neglected regardless of the properties of the particle. The second term in brackets can be discarded if $a|\vec{u}|/(D\beta)\ll 1$, where $|\vec{u}|$ denotes the velocity in the direction of the electric field. In our experiments, both terms inside the curved backets are non-negligible. A detailed theoretical analysis using the canonical model can be found in the paper by Schnitzer \textit{et al.} \cite{Schnitzer2012}.

\section{Coupled Stokes Poisson Nernst-Planck simulation}
\begin{figure}[h]
\includegraphics{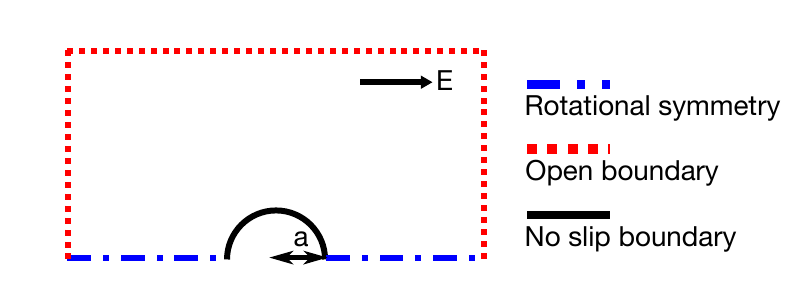}
\caption{\label{fig:domain} The boundary conditions for SPNP simulations (not drawn to scale).}
\end{figure}

We used the coupled Stokes Poisson Nernst-Planck (SPNP) simulation as follows.
The Poisson's equation relates electric potential distribution $\psi$ and volumetric free charge density,
\begin{equation}
\label{eq:CS1}
\epsilon \nabla^2 \psi = -F\sum z_i c_i,
\end{equation}
where $\epsilon$, $F$, $z_i$, and $c_i$ are permittivity of water, Faraday constant, valence, and ion concentration, respectively.
The steady-state Nernst-Planck equation for each ion is
\begin{equation}
\label{eq:CS2}
\nabla \cdot \left(D_i\nabla c_i + z_i \mu_i c_i F \nabla \psi \right) = \vec u \cdot \nabla c_i,
\end{equation}
where $D_i$, $\mu_i$, and $\vec u$ are the diffusion coefficient, mobility of ion, and fluid velocity vector, respectively.
$D_i$ and $\mu_i$ are related by Nernst-Eisntein relation as $\mu_i = D_i/RT$.
The steady state Stokes and continuity equations are
\begin{equation}
\label{eq:CS3}
\eta \nabla^2 \vec u - \nabla p - F \sum z_i c_i \nabla \psi = 0,\quad \nabla \cdot \vec u = 0.
\end{equation}
These equations are solved simultaneously with the finite element method (COMSOL Multiphysics ver.~4.4) in the 2D axisymmetric domain shown in Fig.~\ref{fig:domain}.
A sphere is fixed at the center of the domain. 
The surface of the particle has a constant surface charge density $\sigma$, which is related to the surrounding potential as $-\epsilon \textbf{n}\cdot \nabla \psi = \sigma$, where $\textbf{n}$ is the unit outward normal vector on the surface of the sphere.
The mobility is calculated as the negative of the electroosmotic flow velocity far away from the particle surface.
The physical parameters used for simulations are, temperature: \SI{298.15}{\kelvin}, relative permittivity of water: 80, viscosity of water: \SI{0.889}{\milli\Pa\cdot \second}, diffusion coefficients of ions K$^+$ and Cl$^-$: \SI{1.957}{\times 10^{-9} \metre^2 /\second} and \SI{2.032}{\times 10^{-9} \metre^2 /\second}, respectively.

\section{Asymptotic model of electrophoretic mobility in a low field}
Here we briefly describe how we estimated surface charge densities with the asymptotic model of electrophoretic mobility with relaxation effect by Chen and Keh~\cite{Chen1992}.

The surface potential $\psi_0$ and surface charge density $\sigma$ under zero external field are related with the Grahame equation,
\begin{equation}
\label{eq:grahame}
\sigma = 2\epsilon\kappa \phi_{\textrm{th}} \sinh\left(\frac{\psi_0}{2\phi_{\textrm{th}}}\right).
\end{equation}

The electrophoretic mobility with concentration polarization is calculated using~\cite{Chen1992},
\begin{equation}
\label{eq:mu_ep}
\mu_{\textrm{ep}}= \frac{\epsilon \zeta}{3 \eta} \left[2 + C_1 + C_2 + \left(C_1 - C_2\right)\frac{1}{\bar{\zeta}}\ln \cosh \bar{\zeta}\right],
\end{equation}
where $C_{1, 2}$ are the coefficients of $\kappa a$ and $\zeta$, and $C_{1,2} = 1/2$ in the limit $(\kappa a)^{-1} e^{\left|\zeta\right|/2\phi_{\textrm{th}}} \to 0$.
$\bar{\zeta}$ is described as $\bar{\zeta} = \frac{ze\zeta}{4kT}$.
Eq.~(\ref{eq:mu_ep}) is fitted to the experimental data points in Fig.~2 with least square curve fitting.
The surface charge densities were calculated by assuming $\zeta = \psi_0$.

The terms $C_1$ and $C_2$ are given as,
\begin{subequations}
\begin{equation}
C_1 = \left(\frac{1}{2} + \frac{1}{2}\beta_{22} - \frac{3}{2}\beta_{12} - \beta_{11} + \beta_{12}\beta_{21} - \beta_{11}\beta_{22} \right)/\Delta,
\end{equation}
\begin{equation}
C_2 = \left(\frac{1}{2} + \frac{1}{2}\beta_{11} - \frac{3}{2}\beta_{21} - \beta_{22} + \beta_{12}\beta_{21} - \beta_{11}\beta_{22} \right)/\Delta,
\end{equation}
\end{subequations}
where $\Delta$ and $\beta_{ij}$ are
\begin{equation}
\Delta = 1 + \beta_{11} + \beta_{22} + \beta_{11}\beta_{22} - \beta_{12}\beta_{21},
\end{equation}
\begin{subequations}
\begin{equation}
 \beta_{11} = \frac{1}{\kappa a}\left[4\left(1+\frac{3f_1}{z^2}\right)\exp{\left(\bar{\zeta}\right)}\sinh{\bar{\zeta}} - \frac{12f_1}{z^2}\left(\bar{\zeta} +\ln\cosh\bar{\zeta} \right) \right],
\end{equation}
\begin{equation}
\beta_{12} = \frac{1}{\kappa a}\left(\frac{12f_1}{z^2}\right)\ln\cosh\bar{\zeta},
\end{equation}
\begin{equation}
\beta_{21} = \frac{1}{\kappa a}\left(\frac{12f_2}{z^2}\right)\ln\cosh\bar{\zeta},
\end{equation}
\begin{equation}
\beta_{22} = \frac{1}{\kappa a}\left[-4\left(1+\frac{3f_2}{z^2}\right)\exp\left(-\bar{\zeta}\right)\sinh{\bar{\zeta}} + \frac{12f_2}{z^2}\left(\bar{\zeta} -\ln\cosh\bar{\zeta} \right) \right].
\end{equation}  
\end{subequations}
The parameter $f_1$, $f_2$ and $z$ were set $f_1 =f_2 = 0.2$ and $z = 1$, as suggested for KCl in~\cite{Chen1992, Chiu2016}.

\section{Nonlinear term in Schnitzer and Yariv's model}
Here we describe the Schnitzer and Yariv's model of nonlinear electrophoresis for small Dukhin numbers and arbitrary field strengths~\cite{Schnitzer2014}.
For small P\'eclet numbers, the electrophoretic velocity $v_{\textrm{ep}}$ is calculated as, 
\begin{equation}
\label{eq:schnitzer}
v_{\textrm{ep}} = \left(\frac{\epsilon {\phi_\textrm{th}}^2}{\eta a} \right)\left(\tilde{\zeta} \beta + \textrm{Du} u_1 + \dots \right),
\end{equation}
where $\tilde{\zeta}$ is a dimensionless zeta potential $\tilde{\zeta} = \zeta/\phi_\textrm{th}$. $\textrm{Du}$ is defined as, $\textrm{Du} = \left(1+2\alpha^- \right)\tilde{\sigma}/\left(\kappa a\right)$, where $\alpha^-$ is the dimensionless ionic drag coefficient of counterion (not negative ion), $\alpha^- = \epsilon {\phi_\textrm{th}}^2/(\eta D^{-})$, and $\tilde{\sigma}$ is the dimensionless surface charge density, $\tilde{\sigma} = \sigma/\left( \epsilon \kappa \phi_\textrm{th} \right)$ . 
The dimensionless term $u_1$ is given as,
\begin{equation}
\label{eq:schnitzer_u1}
u_1 \sim -\left[ 4\ln \cosh \left(\frac{\tilde{\zeta}}{4}\right) + \tilde{\zeta}\right]\beta + \frac{2}{21}\beta^3.
\end{equation}
Thus, the nonlinear velocity term $v_\textrm{nl}$ is,
\begin{equation}
\label{eq:schnitzer_nl}
v_{\textrm{nl}} \sim \left(\frac{\epsilon {\phi_\textrm{th}}^2}{\eta a} \right) \textrm{Du} \frac{2}{21}\beta^3.
\end{equation}
The cubic electrophoretic mobility $\mu_{\textrm{ep}}^{(3)}$ ($= v_{\textrm{nl}}/E^3$) is,
\begin{equation}
\label{eq:schnitzer_mu_3}
\mu_{\textrm{ep}}^{(3)} \sim \frac{\epsilon\left(1+2\alpha^- \right)}{21\eta\phi_\textrm{th} z e} \frac{a \sigma}{c N_A},
\end{equation}
where $N_a$ is the Avogadro constant, and $c$ is the concentration (mol/m$^3$).



%